\documentclass[preprint,12pt]{elsarticle}

\usepackage{xcolor}
\usepackage{graphicx}

\usepackage{hyperref}
\usepackage{amssymb}
\usepackage{amsmath}
\usepackage{microtype}
\usepackage{cleveref}

\usepackage[most]{tcolorbox}

\usepackage{listings}

\definecolor{white}{rgb}{1,1,1}
\definecolor{mygreen}{rgb}{0,0.4,0}
\definecolor{light_gray}{rgb}{0.97,0.97,0.97}
\definecolor{mykey}{rgb}{0.117,0.403,0.713}

\tcbuselibrary{listings}
\newlength\inwd
\newcounter{ipythcntr}
\renewcommand{\theipythcntr}{\texttt{[\arabic{ipythcntr}]}}

\newtcblisting{pyin}[1][]{%
  sharp corners,
  enlarge left by=\inwd,
  width=\linewidth-\inwd,
  enhanced,
  boxrule=0pt,
  colback=light_gray,
  listing only,
  top=0pt,
  bottom=0pt,
  overlay={
    \node[
      anchor=north east,
      text width=\inwd,
      font=\footnotesize\ttfamily\color{mykey},
      inner ysep=2mm,
      inner xsep=0pt,
      outer sep=0pt
      ] 
      at (frame.north west)
      {\refstepcounter{ipythcntr}\label{#1}In \theipythcntr:};
  }
  listing engine=listing,
  listing options={
    aboveskip=1pt,
    belowskip=1pt,
    basicstyle=\footnotesize\ttfamily,
    language=Python,
    keywordstyle=\color{mykey},
    showstringspaces=false,
    stringstyle=\color{mygreen}
  },
}
\newtcblisting{pyprint}{
  sharp corners,
  enlarge left by=\inwd,
  width=\linewidth-\inwd,
  enhanced,
  boxrule=0pt,
  colback=white,
  listing only,
  top=0pt,
  bottom=0pt,
  overlay={
    \node[
      anchor=north east,
      text width=\inwd,
      font=\footnotesize\ttfamily\color{mykey},
      inner ysep=2mm,
      inner xsep=0pt,
      outer sep=0pt
      ] 
      at (frame.north west)
      {};
  }
  listing engine=listing,
  listing options={
      aboveskip=1pt,
      belowskip=1pt,
      basicstyle=\footnotesize\ttfamily,
      language=Python,
      keywordstyle=\color{mykey},
      showstringspaces=false,
      stringstyle=\color{mygreen}
    },
}

\hypersetup{%
pdfborder = {0 0 0},
bookmarksdepth = section,
citecolor = DarkGreen,
linkcolor = DarkBlue,
urlcolor = DarkGreen,
bookmarksnumbered = true
}%

\biboptions{numbers,square,sort&compress}

\newcounter{bla}

\makeatletter
\def\ps@pprintTitle{%
  \let\@oddhead\@empty
  \def\@oddhead{\hfill MITP-22-052, MS-TP-22-31}%
  \let\@evenhead\@empty
  \let\@oddfoot\@empty
  \let\@evenfoot\@oddfoot
}
\makeatother

\begin{document}

\begin{frontmatter}

\title{pyerrors: a python framework for error analysis of \\Monte Carlo data}

\author[a]{Fabian Joswig\corref{author}}
\author[b,c]{Simon Kuberski}
\author[d]{Justus T. Kuhlmann}
\author[d]{Jan Neuendorf}

\cortext[author] {Corresponding author.\\\textit{E-mail address:} fabian.joswig@ed.ac.uk}
\address[a]{Higgs Centre for Theoretical Physics, School of Physics and Astronomy,\\
The University of Edinburgh, Edinburgh EH9 3FD, UK}
\address[b]{Helmholtz-Institut Mainz, Johannes Gutenberg-Universität Mainz,\\
Staudingerweg 18, 55128 Mainz, Germany}
\address[c]{GSI Helmholtzzentrum für Schwerionenforschung,\\
Planckstraße 1, 64291 Darmstadt, Germany}
\address[d]{Institut für Theoretische Physik, Westfälische Wilhelms-Universität Münster,
Wilhelm-Klemm-Straße 9, 48149 Münster, Germany}

\begin{abstract}
We present the \texttt{pyerrors} \texttt{python} package for statistical error analysis of Monte Carlo data. Linear error propagation using automatic differentiation in an object oriented framework is combined with the $\Gamma$-method for a reliable estimation of autocorrelation times. Data from different sources can easily be combined, keeping the information on the origin of error components intact throughout the analysis. \texttt{pyerrors} can be smoothly integrated into the existing scientific \texttt{python} ecosystem which allows for efficient and compact analyses.
\end{abstract}


\end{frontmatter}

\newpage

\section{Introduction}
The Standard Model of particle physics (SM) is very successful in describing the basic constituents of matter and their interactions but falls short to explain a number of observed phenomena such as the matter-antimatter asymmetry of the universe or the existence of dark matter. The search for physics beyond the Standard Model relies on highly precise theoretical calculations of SM observables. Lattice QCD is the only known ab-initio approach to compute these observables in the low energy, hadronic regime of quantum chromodynamics (QCD), describing the strong interactions in the SM. The method is based on a Markov Chain Monte Carlo sampling of the QCD action in Euclidean space, formulated via the path integral formalism.

In recent years, lattice QCD calculations have become a precision tool such that they have a relevant impact on phenomenology and the search for beyond the SM theories, see Reference \cite{FlavourLatticeAveragingGroupFLAG:2021npn} for a recent review of lattice results. Prime examples are the computation of the strong coupling constant of QCD and of hadronic contributions to the anomalous magnetic moment of the muon \cite{Colangelo:2022jxc}, where recent studies achieve precision similar to data-driven methods. The main contribution to the systematic uncertainties of these calculations and therefore the main obstacle for obtaining even higher precision is due to the extrapolation to the continuum limit of the theory.

When simulating boxes with ever-increasing resolution, thereby lifting the ultraviolet cutoff, state of the art algorithms experience critical slowing down. This results in a significant increase of the autocorrelations towards the continuum limit \cite{Schaefer:2010hu}, which are inherent in Markov Chain Monte Carlo (MCMC) methods. Thus, the analysis of data stemming from MCMC methods requires a careful treatment of these autocorrelations. Otherwise statistical errors might be underestimated, putting the significance of results in question.

The $\Gamma$-method as introduced in \cite{Madras1988,Wolff:2003sm} has been shown to decrease systematic errors in the treatment of autocorrelation in comparison to widely used techniques such as binned Jackknife \cite{Quenouille1956,Tukey1958} or Bootstrap resampling \cite{Efron1979}, especially in the presence of critical slowing down \cite{Schaefer:2010hu}. Whereas the first publicly available implementations \cite{Wolff:2003sm, Schaefer:2010hu, DePalma:2017lww} do not allow to combine observables from simulations with different parameters, this is possible in the implementations of \cite{Ramos:2018vgu, Ramos:2020scv} written in \texttt{Fortran} and \texttt{Julia}.\footnote{We note that the \texttt{MATLAB} package \texttt{dobs} \cite{dobs} that is used within the ALPHA collaboration and predates the other implementations as well as the \texttt{pyobs} \cite{pyobs} package provide similar features.}

In this article, we present the \texttt{pyerrors} package that provides a framework to estimate and propagate uncertainties of Monte Carlo data in \texttt{python}. The guiding design principle of \texttt{pyerrors} is to provide a new data type, \texttt{Obs}, which integrates into the existing scientific \texttt{python} ecosystem as smoothly as possible. \texttt{Obs} can be the argument of most \texttt{numpy} \cite{harris2020array} functions, thus allowing for an easy and fast implementation of analyses using the $\Gamma$-method. The linear propagation of uncertainties is performed to machine precision via automatic differentiation \cite{Ramos:2018vgu, maclaurin2015autograd}. \texttt{pyerrors} has already been used in a set of scientific publications \cite{Heitger:2020mkp,Heitger:2020zaq,Heitger:2021apz,Heitger:2021bmg,Blossier:2021xvl,Ce:2022kxy}. While this article is focused on data analysis in the context of lattice QCD, the \texttt{pyerrors} package introduced here may also be applied to other kinds of correlated data.

This paper is organized as follows. In \Cref{s:MCMC} we briefly summarize how error propagation and error estimation are carried out in the $\Gamma$-method approach in combination with automatic differentiation. We proceed with a description of our implementation of these concepts within \texttt{pyerrors} in \Cref{s:implementation}. In \Cref{s:examples}, we collect examples to illustrate the basic features of the package. In addition to the core routines that are used for the error computation, a number of other features has been implemented to facilitate frequently encountered tasks such as the computation of effective masses from correlation functions, fits to data with uncertainties and the solution of a Generalized Eigenvalue Problem (GEVP) \cite{wilson:gevp,Michael:1982gb,Luscher:1990ck,Blossier:2009kd}. Some of these routines are presented in \Cref{s:examples}. Users are welcome to contribute to extending the functionality of \texttt{pyerrors} via the repository \url{https://github.com/fjosw/pyerrors}.

\section{Analysis of Markov chain Monte Carlo data \label{s:MCMC}}

\subsection{Description of the problem}
Data that is obtained from a Markov Chain Monte Carlo simulation exhibits autocorrelation -- subsequent measurements are not independent from each other because the underlying distributions evolve by sequential updates. Such autocorrelations in the data reduce the effective number of independent measurements. To obtain an accurate estimate of the statistical uncertainty of an observable, the autocorrelation has to be taken into account appropriately.

In the context of lattice QCD, the correct treatment of autocorrelation becomes more and more important as the field enters the precision era, where sub-percent precision is reached for phenomenologically important quantities computed at fine resolutions below $0.05\,$fm. Due to critical slowing down \cite{Schaefer:2010hu}, a significant increase of autocorrelation times is expected and observed towards the continuum limit for all algorithms that are currently being used for large-scale simulations. Sufficiently long Monte Carlo chains and a precise treatment of autocorrelations are vital to ensure the correctness of error estimates.

Traditionally, the estimation of uncertainties of Monte Carlo observables is performed using resampling techniques such as Jackknife or Bootstrap resampling. In the presence of autocorrelation, primary data is binned to blocks in Monte Carlo time that are large compared to the autocorrelation time of the observable.\footnote{See \Cref{ssec:error_estimation} for the definitions of the autocorrelation function and the integrated autocorrelation time of a Monte Carlo observable.} This may be problematic if long-ranged autocorrelation is only present in complicated derived observables and not visible in primary data that is used to determine the appropriate bin size or if bin sizes are large compared to the total number of measurements. In the latter case, only a small number of bins may be used within the resampling techniques, thus enlarging the error of the error estimation itself.

The aforementioned problems have been addressed in Ref.~\cite{Wolff:2003sm}, where the so-called $\Gamma$-method has been introduced as a tool to investigate the autocorrelation function $\Gamma(t)$ of primary and secondary observables and to use it to correctly scale the standard error. When combining data from different sources, e.g., Monte Carlo simulations with different parameters, the treatment within the context of the $\Gamma$-method furthermore allows to cleanly separate these sources of uncertainty in arbitrarily complicated derived observables. While this can also be done within resampling approaches, such separation is intrinsic to the $\Gamma$-method. Therefore, the management of meta-data and self-contained long-time storage in compliance with the FAIR principles for scientific data management and stewardship (Findability, Accessibility, Interoperability, and Reuse of digital assets) \cite{Wilkinson2016} is facilitated.

\subsection{Error propagation}
The general route in Monte Carlo calculations in order to obtain estimators for observables $A_\alpha$ is to generate a sufficiently large number $N$ of samples $a_\alpha^i$ which obey the given probability distribution and take the mean value
\begin{align}
	\bar{a}_\alpha=\frac{1}{N}\sum_{i=1}^{N}a_\alpha^i\,,\quad A_\alpha=\lim_{N\to\infty}\bar{a}_\alpha\,. \label{eq:montecarlo_mean}
\end{align}
An observable is furthermore characterized by the fluctuations
\begin{align}
\delta^i_\alpha=a_\alpha^i -\bar{a}_\alpha^{\phantom{i}}\,,
\end{align}
which contain all relevant information for the computation of the statistical error, as the variance and its more general variant, the autocorrelation function, are invariant under constant shifts of the data.
One is often interested in a derived observable which can be an arbitrary function of any number of primary observables
\begin{align}
    f(A_\alpha)=f(\bar{a}_\alpha)\,.
\end{align}
The error propagation we use for the consequent error estimation is based on a first-order Taylor series expansion as detailed in Ref.~\cite{Wolff:2003sm}
\begin{align}
\delta_f^i=\sum_{\alpha}\bar{f}_\alpha\delta_\alpha^i+\dots\,, \label{eq:delta_resampling}
\end{align}
where the derivatives are defined as
\begin{align}
\bar{f}_\alpha=\frac{\partial f}{\partial A_\alpha}\bigg|_{A_\alpha=\bar{a}_\alpha}\,.\label{eq:derivatives}
\end{align}
A derived observable is then again characterized by its mean $f(\bar{a}_\alpha)$ and its fluctuations $\delta_f^i$ and the distinction between primary and derived observables becomes indistinguishable \cite{Virotta2012Critical}. The numerical estimation of the derivative in \cref{eq:derivatives} as proposed in Ref.~\cite{Wolff:2003sm} may be replaced by the derivative obtained from automatic differentiation \cite{Ramos:2018vgu}, making the linear error propagation exact to machine precision.\footnote{This type of error propagation strategy also works for the outputs of iterative algorithms like non-linear least squares or root finding by deriving the required derivatives for \cref{eq:derivatives} as described in detail in Ref.~\cite{Ramos:2018vgu}.}

\subsection{Error estimation}
\label{ssec:error_estimation}
In order to estimate the statistical Monte Carlo error of any primary or derived observable one can write the estimator for the autocorrelation function in terms of the fluctuations $\delta_f^i$ as
\begin{align}
\Gamma_f(t) = \frac{1}{N-t}\sum_{i=1}^{N-t}\delta_f^i\delta_f^{i+t}\,.
\label{eq:gamma_function}
\end{align}
which for $N\to\infty$ becomes the exact autocorrelation function.
Note that $\Gamma_f(0)=\operatorname{var}(\delta_f^i)$.
The autocorrelation function is expected to exhibit an exponential decay for large $t$ according to 
\begin{align}
\frac{\Gamma_f(t)}{\Gamma_f(0)}\sim \exp\Bigg( -\frac{t}{\tau_{\mathrm{exp}}} \Bigg)\,,
\end{align} 
where the exponential autocorrelation time $\tau_{\mathrm{exp}}$ is the largest autocorrelation time in the system. The corresponding integrated autocorrelation time for observable $f$ is estimated via \cite{Madras1988}
\begin{align}
\tau_{\mathrm{int},f}(W)=\frac{1}{2}+\sum_{t=1}^{W}\frac{\Gamma_f(t)}{\Gamma_f(0)}\,.\label{eq:tauint_gamma}
\end{align}
One practical problem is to find an appropriate window $W$ for which all relevant information about the autocorrelation is captured in $\tau_{\mathrm{int},f}$ while at the same time the noise contribution for large values of $t$ does not overwhelm the signal \cite{Madras1988}.
In Ref.~\cite{Wolff:2003sm} an automatic windowing procedure was suggested which minimizes the sum of statistical and systematic errors based on the hypothesis that $\tau_\mathrm{exp} \sim S \tau_{\mathrm{int},f}$ with some factor $S$.

The failure to detect a coupling to the slowest modes in the system in practical applications, e.g., because the Monte Carlo chain is not long enough, might lead to an underestimation of $\tau_{\mathrm{int},f}$. In Ref.~\cite{Schaefer:2010hu} an extension to the estimation of the integrated autocorrelation time was suggested that accounts for the effect of $\tau_\mathrm{exp}$ in the error estimate. The idea is to attach a tail to the autocorrelation function,
\begin{align}
\tau_{\mathrm{int},f}(W)=\frac{1}{2}+\sum_{t=1}^{W}\frac{\Gamma_f(t)}{\Gamma_f(0)}+\tau_\mathrm{exp}\bigg|\frac{\Gamma_f(W+1)}{\Gamma_f(0)}\bigg|\,,\label{eq:tail}
\end{align}
based on prior knowledge of $\tau_\mathrm{exp}$. Also in eq.~(\ref{eq:tail}) one has to find an appropriate summation window $W$. Instead of using the automatic windowing procedure of Ref.~\cite{Wolff:2003sm}, it is custom to use the first value of $W$ for which $\rho(t)=\Gamma_f(t)/\Gamma_f(0)$ is zero within its error. The error of $\rho(t)$ can be approximated by \cite{Luscher:2004pav}\footnote{This approximation is not guaranteed to be correct in all cases, see e.g., \cite{Virotta2012Critical}.}
\begin{align}
\big( \delta\rho(t) \big)^2\approx\frac{1}{N}\sum_{m=1}^{W}\big( \rho(m+t)+\rho(m-t)-2\rho(m)\rho(t) \big)^2\,. \label{eq:err_of_err}
\end{align}
The final error estimate based on the integrated autocorrelation time $\tau_{\mathrm{int},f}$ can be computed via
\begin{align}
    \sigma_f=\sqrt{2\tau_{\mathrm{int},f}\frac{\operatorname{var}(\delta_f^i)}{N}}\,,\label{eq:monte_carlo_error}
\end{align}
which reduces to the standard error in the uncorrelated case for which $\tau_{\mathrm{int},f}=0.5$.

\subsection{Comparison to other error estimation techniques}

In this section we compare the error estimation strategy which we outlined above to two of the most used techniques in the lattice field theory community -- Jackknife \cite{Quenouille1956,Tukey1958} and Bootstrap \cite{Efron1979} resampling.
All three approaches have in common that they are associative, so analyses can be split up in multiple smaller steps in order to obtain more complicated derived observables.
In contrast to Jackknife or Bootstrap resampling the $\Gamma$-method requires the application of non-linear functions in this process only once. In turn, one requires the derivative of the result with respect to every input observable. For a large number of samples and complicated functions the computation of the derivatives can be cheaper than evaluating the function for every sample as one would do it in a traditional resampling approach \cite{Ramos:2018vgu}.\footnote{In the context of variational methods like the GEVP, this means that the eigenvalue problem has to be solved only once and the ordering of the states therefore is unique. This may result in an enhanced stability of the subsequent error estimate, compared to resampling techniques.}

A benefit of the $\Gamma$-method approach is that the systematic error in the estimation of the autocorrelation that comes with the truncation of the sum in \cref{eq:tauint_gamma} decreases asymptotically as $\exp(-W/\tau_\mathrm{exp})$. The systematic error from a binning procedure on the other hand only decreases asymptotically as $\tau_\mathrm{exp}/B$ where $B$ is the bin size as demonstrated in Ref.~\cite{Wolff:2003sm}. For sufficiently many samples $N$ the ratio of the systematic to the statistical error becomes negligible for the $\Gamma$-method whereas the ratio stays constant in case of a binning analysis.

Another more practical advantage of the $\Gamma$-method error analysis strategy is that all the required information for a dedicated autocorrelation analysis is always available for any derived observable by construction. In the case of binning analyses combined with Jackknife or Bootstrap resampling one has to make some choice for the bin sizes of each ensemble, possibly determined from a series of standard observables. This can be dangerous in cases where very long autocorrelation times arise in specific derived observables but were not visible in the primary observables that contribute to it (for examples of such cases see Ref.~\cite{Ramos:2018vgu}). A dedicated study of the evolution of covariances with the bin size including an extrapolation to infinite bin size, as outlined in Appendix
D of Ref.~\cite{RQCD:2022xux}, may resolve this problem in binning analyses.

One advantage of Bootstrap resampling over both Jackknife resampling and the linear error propagation we advocate is that it can capture non-linear distributions and thus non-symmetric confidence intervals. However, since a fundamental assumption of Monte Carlo simulations is the central limit theorem, a linear approximation should reproduce the correct distribution in the infinite sample limit.

\section{Implementation \label{s:implementation}}

At the core of the \texttt{pyerrors} implementation stands the \texttt{Obs} class which provides the user with a new \texttt{python} data type for Monte Carlo observables. An \texttt{Obs} object is initialized with a set of Monte Carlo samples and a corresponding string which serves as unique identifier for the Monte Carlo ensemble.\footnote{The error propagation also works for multiple independent Monte Carlo streams with identical parameters, often referred to as replica.} In addition the user can explicitly provide the set of configurations on which the samples were obtained also allowing for irregular or gapped Monte Carlo histories. We refer the reader to \ref{app:irreg} for an explanation of our implementation of irregular Monte Carlo chains.

Elementary mathematical operations are overloaded for the \texttt{Obs} class. The user can, e.g., multiply an \texttt{Obs} object with an integer, a floating point number or another \texttt{Obs} object and the underlying error propagation is taken care of intrinsically following \cref{eq:delta_resampling}, where the required derivatives are obtained via automatic differentiation \cite{Ramos:2018vgu}. Overloading also works for most \texttt{numpy} functions such as \texttt{log} or \texttt{exp} allowing the user to design an analysis without having to think about cross- or autocorrelations. \texttt{Obs} objects can also be arguments of iterative algorithms like non-linear least squares minimization or root finding. In order to obtain the required derivatives for arbitrary functions we use the \texttt{autograd} package \cite{maclaurin2015autograd}.
For complex valued observables we also provide a \texttt{CObs} class in analogy to the \texttt{python} \texttt{complex} type.

The raw results of Markov Chain Monte Carlo calculations in lattice QCD are often not single observables but Euclidean time dependent correlation functions of two (or more) operators.
To conveniently handle correlation functions in analyses we provide the \texttt{Corr} class which can be considered an array class for \texttt{Obs} objects. In addition, the class has a set of regularly used methods for the manipulation of correlator objects. The user can, e.g., obtain the effective mass or discrete derivatives of a correlator by calling a class method which returns a new \texttt{Corr} object.
The \texttt{Corr} class can also handle matrix valued correlation functions and may be used to solve a GEVP \cite{wilson:gevp,Michael:1982gb,Luscher:1990ck,Blossier:2009kd}.
Having access to data types which map onto the relevant concepts simplifies analyses, makes them more readable and less prone to error. The \texttt{Corr} class for example eliminates the need for any explicit loops over the temporal argument of the correlation function.

\subsection{Error estimation}
After arriving at a desired arbitrary function of primary observables (which can be obtained in multiple steps as explained in the previous sections) the error of this quantity can be estimated by calling the method \texttt{gamma\_method} of the \texttt{Obs} object. We implemented both, the automatic windowing procedure of Ref.~\cite{Wolff:2003sm} and the extended variant proposed in Ref.~\cite{Schaefer:2010hu}.
The computationally most expensive part in estimating the Monte Carlo error along the lines described in \Cref{ssec:error_estimation} is the computation of the autocorrelation function according to \cref{eq:gamma_function}.\footnote{This only has to be done when the error is needed, i.e., not for intermediate quantities.} The computation can be sped up by making use of the Wiener-Khintchine theorem \cite{Wiener1930,Khintchine1934} and the real fast Fourier transform algorithm ($\operatorname{rfft}$) \cite{Sorensen1987} with computational complexity $\mathrm{O}(N\log N)$.
To reproduce the exact expression of \cref{eq:gamma_function} a linear convolution can be used which requires extending the vector $\delta_f$ by $t_\mathrm{max}$ zeros to ensure that there are no problems with the periodicity assumed by the $\operatorname{rfft}$ routine,
\begin{align}
({\delta}^\prime)_f^{i} &= \begin{cases}
{\delta}_f^i & \, \,\phantom{N+}1 \leq i \leq N \\
0 & N+1 \leq i \leq N+t_\mathrm{max} \label{eq:delta_fft}
\end{cases}\,,\\
\Gamma_f(t) &= \frac{1}{N-t}\Big[\operatorname{rfft}^{-1}\big(|\operatorname{rfft}\big(({\delta}^\prime)_f\big)|^2\big)\Big](t)\,. \label{eq:fft}
\end{align}
In practice we compute the autocorrelation function up to $t_\mathrm{max}=N/2$.

The error estimation for a single Monte Carlo ensemble can be easily extended to analyses to which multiple, independent Monte Carlo ensembles contribute. Within a \texttt{pyerrors} \texttt{Obs} we keep track of the fluctuation of each ensemble independently and uniquely identify it with a string. The error contributions from independent ensembles can then be estimated individually via \cref{eq:monte_carlo_error} and added in quadrature. This procedure allows to cleanly separate error contributions from independent ensembles and prevents contamination with spurious correlation which might arise in combined Bootstrap analyses.

Many analyses make use of information that is provided as external input. Examples for such external information in the context of lattice field theory are experimentally determined quantities such as meson masses or decay constants which are used to calibrate the theory.
The treatment of (correlated) external observables via Gaussian error propagation is straight-forward in the context of the $\Gamma$-method with automatic differentiation. The knowledge of the covariance matrix $\mathcal{C}$ for a set of external observables $p_i$ together with the Jacobian matrix for the computation of a secondary quantity $\mathcal{O} = f(p_i)$ which is defined via 
\begin{align}
    J = \frac{\partial f}{\partial p_i}
\end{align}
may be used to determine the contribution of the $p_i$ to the total error of $\mathcal{O}$ via
\begin{align}
    s_p = \sqrt{J^\mathrm{T}\mathcal{C} J}\,.
\end{align}
This computation is exact in the sense of Gaussian error propagation. The information on the derivative of any secondary observable with respect to $p_i$ as well as on the covariance between the $p_i$ may be propagated through the entire analysis. The clean separation of Monte Carlo data from different independent ensembles as well as correlated constants as external input allows one to monitor the individual contributions to the squared error.

\section{Examples \label{s:examples}}
In this section we provide a few minimal examples which demonstrate the capabilities of the \texttt{pyerrors} framework. For a full account of the functionality see the online documentation \cite{pyerrorsdoc}.
\subsection{Mathematical operations}
An \texttt{Obs} object which carries the information of a single Monte Carlo observable can be constructed by providing two lists, one containing the samples, the other the names of the individual Monte Carlo chains on which the observable has support on
\begin{pyin}[cell:mat_op]
import pyerrors as pe
my_obs = pe.Obs([sample_data], ["A1"])
\end{pyin}
\noindent
Optionally the identifiers of the configurations can be explicitly specified and tracked via the argument \texttt{idl}. For data on every second configuration between 1 and 2000 one could specify
\begin{pyin}[cell:idl]
my_obs = pe.Obs([sample_data], ["A1"],
                idl=[range(1, 2000, 2)])
\end{pyin}
\noindent If the argument \texttt{idl} is not given it is assumed that the first sample corresponds to configuration 1 and that there are no gaps in between configurations.

An \texttt{Obs} object can then be modified with arbitrary elementary operations as well as \texttt{numpy} functions which are overloaded for the \texttt{Obs} class
\begin{pyin}[cell:derived_obs]
import numpy as np
derived_obs = np.log(my_obs ** 2)
\end{pyin}

\noindent
After arriving at the derived observable of interest the error can be estimated via the \texttt{gamma\_method}.
\begin{pyin}[cell:gamma_method]
derived_obs.gamma_method()
print(derived_obs)
\end{pyin}

\begin{pyprint}
0.2611(43)
\end{pyprint}

The default behavior of the \texttt{gamma\_method} is to use the automatic windowing procedure of Ref.~\cite{Wolff:2003sm} with the parameter \texttt{S} $=2.0$. The user can call the method with a different value for \texttt{S} as argument or with an estimate of the exponential autocorrelation time via the argument \texttt{tau\_exp} in order to refine the estimation of the autocorrelation.
If \texttt{S} $=0$ is specified, the autocorrelation analysis is deactivated and the standard error is calculated instead.
\subsection{Correlation functions and the effective mass}

Lattice data which is defined on a set of timeslices can be represented using the \texttt{Corr} class
\begin{pyin}[cell:corr]
corr = pe.Corr(list_of_obs)
\end{pyin}
\noindent
\texttt{Corr} objects support arithmetic operations with numbers, \texttt{Obs} and other \texttt{Corr}s via operator overloading.
They also implement the \texttt{gamma\_method} and other commonly used operations. 
As an example, we can obtain the log effective mass of the correlation function $C(x_0)$
\begin{align}
    am_\mathrm{eff}(x_0)=\log\frac{C(x_0)}{C(x_0+1)}\,,
\end{align}
by calling the method \texttt{m\_eff} which returns another \texttt{Corr} object
\begin{pyin}[cell:m_eff]
am_eff = corr.m_eff(variant='log') 
am_eff.gamma_method()
\end{pyin}
\noindent
The new \texttt{Corr} object \texttt{am\_eff} has the same length as \texttt{corr}. The last entry, which is not defined in this case, is filled with \texttt{None}.
We also implemented other variants of the effective mass, e.g., the cosh effective mass for periodic boundary conditions.

After having identified an appropriate region in which the effective mass appears to form a plateau the corresponding value can be extracted by calling the method \texttt{plateau} with the desired range as argument. The method returns an \texttt{Obs} object and we retain access to all tools for error analysis
\begin{pyin}[cell:plateau]
am = am_eff.plateau(plateau_range=[5, 18]) 
am.gamma_method()
print(am)
\end{pyin}
\begin{pyprint}
0.0539(50)
\end{pyprint}

\subsection{Least square fitting of data}
Suppose we want to fit the correlation function with an exponential decay
$f(t)=Z^2 e^{-amt}$.
We define a fit function with two parameters as follows
\begin{pyin}[cell:fit_function]
import autograd.numpy as anp
def fitf(a, x):
    return a[0] ** 2 * anp.exp(-a[1] * x)
\end{pyin}
\noindent
We can then perform an uncorrelated least squares fit by calling the corresponding class method
\begin{pyin}[cell:fit]
fit_result = corr.fit(fitf, fitrange=[5, 26])
\end{pyin}
\noindent
From this fit result, we can get the values of $Z$ and $am$ and estimate their errors
\begin{pyin}[cell:fit_result]
fit_result.gamma_method()
Z, am_fit = fit_result
print(Z, am_fit)
\end{pyin}
\begin{pyprint}
0.0243(10) 0.0536(46)
\end{pyprint}
\noindent
Once again, the results are \texttt{Obs} objects which can be reused in subsequent analysis steps. \texttt{pyerrors} also supports more complicated fitting procedures like fully correlated fits or fits involving errors in the independent variables via the total least squares method.

\subsection{Error contributions from different sources}
\begin{figure}[ht!]
    \centering
    \includegraphics[scale=0.7]{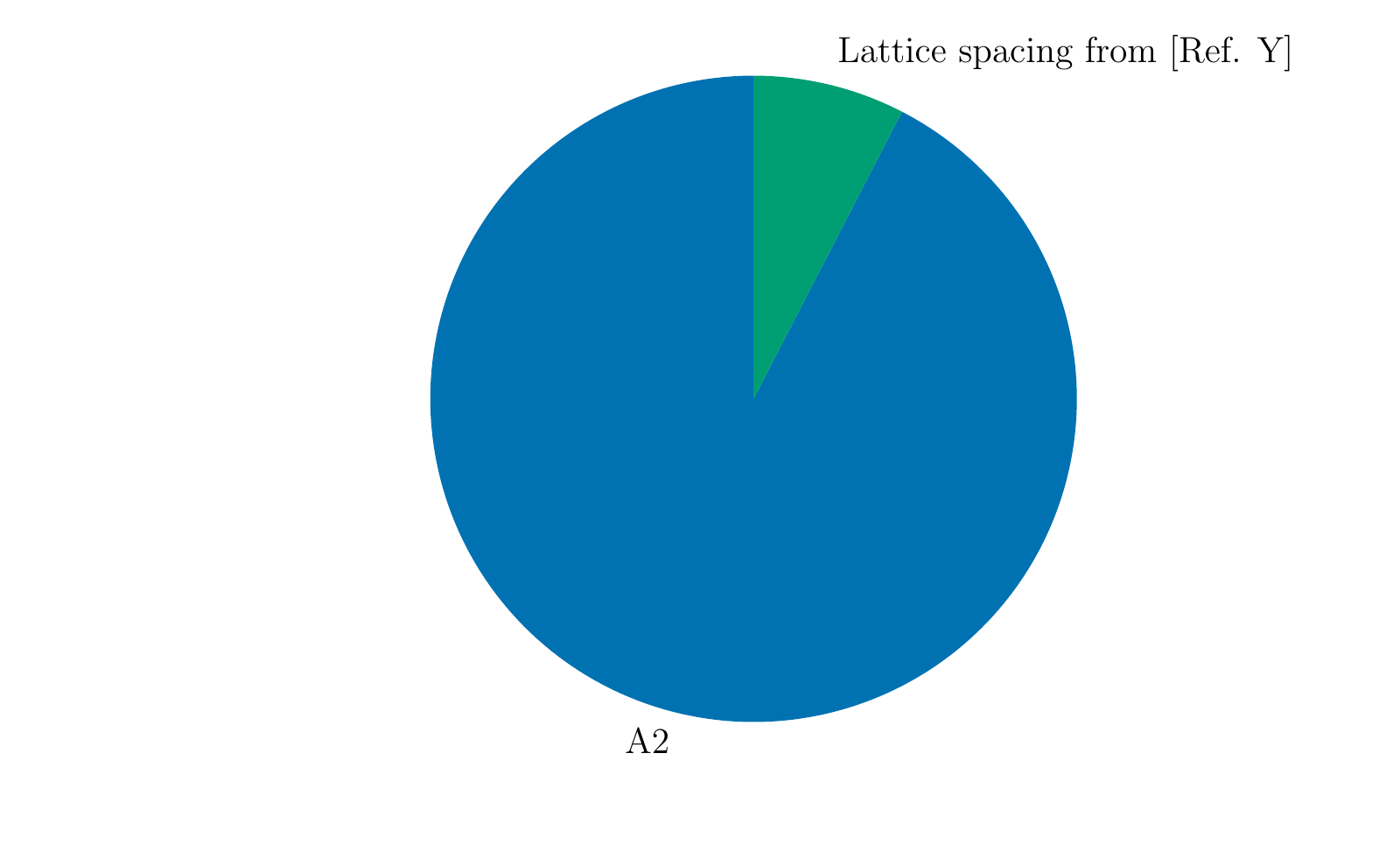}
    \caption{Output of cell \ref{cell:piechart}, visualizing the relative contributions to the squared total error.}
    \label{fig:piechart}
\end{figure}

For quantities with an error not originating from Monte Carlo simulations we can use the \texttt{Covobs} class which is initialized with a value and a squared error.\footnote{While we only need a single value here, the class can also represent a set of quantities with non-zero covariance.} In the following example we convert the previously extracted mass to physical units using a lattice spacing which is known as external input.
\begin{pyin}[cell:cov_obs]
hbar_c = 197.3269804  # MeV*fm
a = pe.cov_Obs(0.0877, 0.0019 ** 2, 
               'Lattice spacing from [Ref. Y]')
m = hbar_c * am / a
\end{pyin}
\noindent
After estimating the error of the mass via the \texttt{gamma\_method} we can display more details about the result by calling the \texttt{details} method
\begin{pyin}[cell:m_result]
m.gamma_method()
m.details()
\end{pyin}
\begin{pyprint}
Result	 1.28799101e+02 +/- 1.01275418e+01 (7.863%)
 Ensemble errors:
 A2 	 9.73554095e+00 +/- 1.72788711e+00
 t_int	 5.90703503e+00 +/- 1.79169621e+00 S = 2.00
 Lattice spacing from [Ref. Y] 	 2.79040242e+00
1000 samples in 2 ensembles:
 Ensemble 'A2' : 1000 configurations (from 1 to 1000)
 Covobs   'Lattice spacing from [Ref. Y]' 
\end{pyprint}
\noindent
When more than one Monte Carlo ensemble or external quantity contribute to the final error we can visualize the relative contributions to the squared error in the following way
\begin{pyin}[cell:piechart]
m.plot_piechart()
\end{pyin}
\noindent
the corresponding output of cell \ref{cell:piechart} is shown in Fig.~\ref{fig:piechart}.

\subsection{The generalized eigenvalue problem}

\texttt{Corr} objects can also be used to represent matrices of observables defined at different time slices of the lattice.
This is very useful, if one wants to solve a GEVP. 
We consider a correlator matrix which consists of symmetric $N\times N$ matrices at every timeslice $t$,
\begin{align}
    C(t)=C_{ij}(t)=\langle O_i(t)O_j^\ast(0) \rangle,
\end{align}
where $O_i,O_j$ are different operators with the same quantum numbers, e.g., operators with different levels of Gaussian smearing applied. 

In order to extract the individual energy levels we can solve the GEVP for a reference time $t_0$ on every timeslice
\begin{align}
    C(t)v_i(t, t_0)=\lambda_i(t,t_0) C(t_0)v_i(t, t_0)\,,\quad v_i(t, t_0)^T C(t_0) v_i(t, t_0)=1\,.
\end{align}
\begin{pyin}[cell:gevp_solution]
evs = C.GEVP(t0=2)
\end{pyin}
\noindent
We can then obtain the corresponding eigenvalues on each timeslice via the relation $C_{i}(t)=v_i(t, t_0)^T C(t)v_i(t, t_0)$ by calling the \texttt{projected} method and calculate the corresponding effective masses in one line
\begin{pyin}[cell:gevp_projection]
am_eff_0 = C.projected(evs[0]).m_eff()
am_eff_0.tag = "Ground state"
am_eff_1 = C.projected(evs[1]).m_eff()
am_eff_1.tag = "First excited state"
\end{pyin}
\noindent

\begin{figure}[ht!]
    \centering
    \includegraphics[scale=0.75]{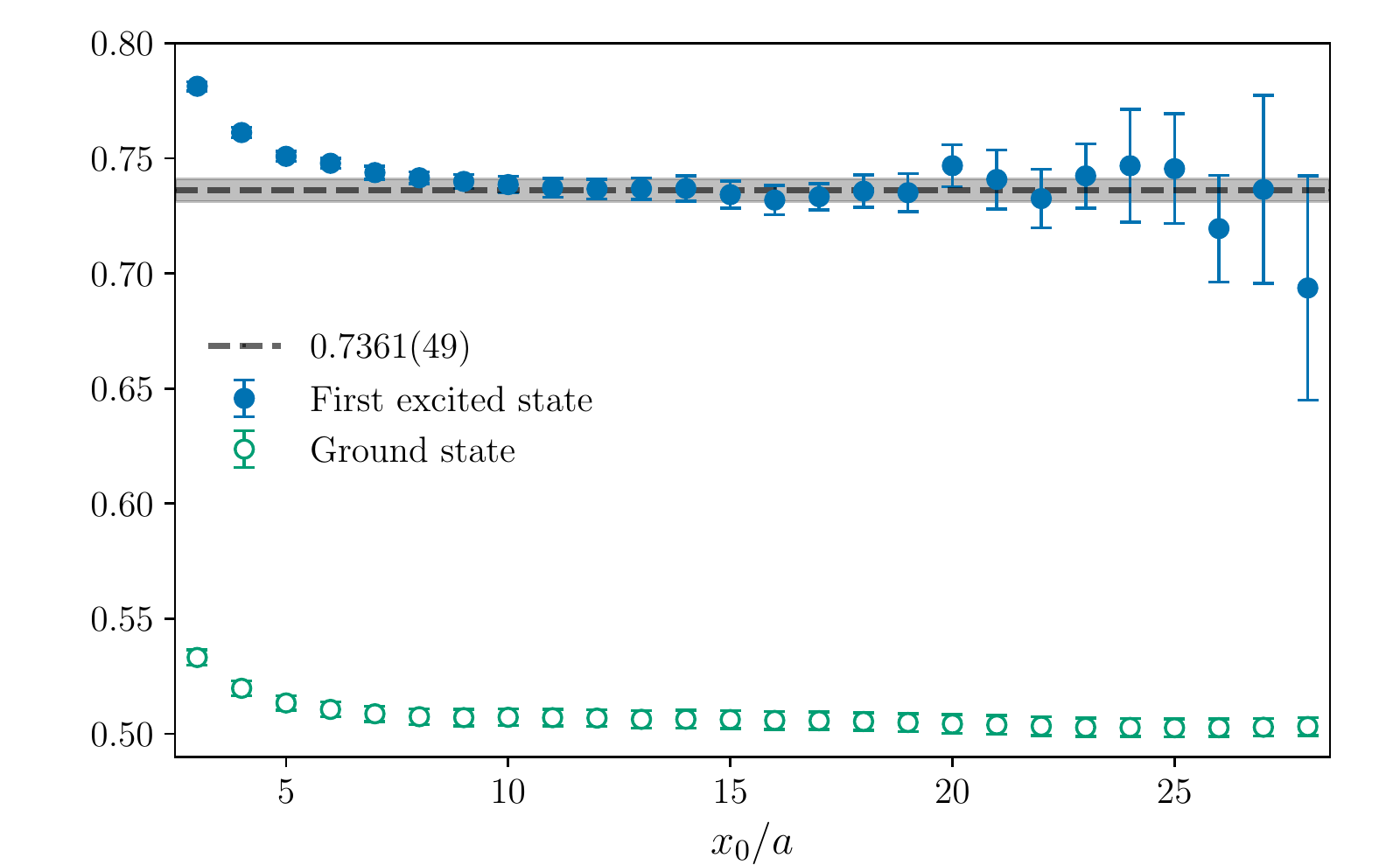}
    \caption{Output of cell \ref{cell:gevp_viz}, visualizing the extraction of the energy of the first excited state via a GEVP.}
    \label{fig:GEVP_m_eff}
\end{figure}

After extracting a plateau value we can visualize the effective masses together with the plateau
\begin{pyin}[cell:gevp_viz]
am1 = am_eff_1.plateau([12, 24])
am1.gamma_method()
am_eff_1.show(x_range=[3, 28], y_range=[0.49, 0.8],
              comp=am_eff_0, plateau=am1)
\end{pyin}
\noindent
The output of cell \ref{cell:gevp_viz} is shown in Fig.~\ref{fig:GEVP_m_eff}.

In the example above the eigenvalues are sorted for every timeslice independently. We also implement additional methods from Ref.~\cite{Fischer:2020bgv} to sort the eigenvectors, if they are independently computed on different timeslices.

\subsection{Data export and storage}
To allow for reusability of stored data, metadata and the data itself have to be processed and stored together in accordance with the FAIR principles \cite{Wilkinson2016}. One vital piece of information, the origin of the data, is propagated throughout an entire analysis with \texttt{pyerrors}.
The export and import of native \texttt{pyerrors} structures such as \texttt{Corr} and \texttt{Obs} or lists and \texttt{numpy.ndarray}s of \texttt{Obs} is implemented via compressed \texttt{JSON} files. In addition to the data, these contain information about the author, the creation date and location and the version of \texttt{pyerrors}. The file content may be further described by a \texttt{python} string, list or dictionary
\begin{pyin}[cell:output]
desc = "Fit to a PP correlation function."
pe.input.json.dump_to_json([Z, am_fit],
                           fname='fit_data', 
                           description=desc)
\end{pyin}
\noindent
All relevant data and metadata can then be restored at any later stage within the \texttt{Obs} objects
\begin{pyin}[cell:input]
Z, am_fit = pe.input.json.load_json('fit_data')
\end{pyin}
\noindent
The \texttt{JSON} serialization also facilitates the storage of \texttt{pyerrors} structures in relational databases.

\section{Conclusions}
\texttt{pyerrors} combines well-known techniques for the estimation and propagation of uncertainties in the context of Monte Carlo data. The framework allows to implement complex analyses in the existing scientific \texttt{python} ecosystem. This enables the user to profit from the known benefits of the $\Gamma$-method without having to care about the propagation of uncertainties, which is performed in the background. While we have explained the features of \texttt{pyerrors} in the context of lattice field theories, the implementation may be helpful for any kind of Monte Carlo data that exhibits autocorrelation.

\section*{Acknowledgments}
We thank the members of the ALPHA collaboration and especially Mattia Bruno, Jochen Heitger, Alberto Ramos and Rainer Sommer for sharing experiences and insights concerning the treatment of statistical uncertainties. We like to thank Rainer Sommer for a critical reading of the manuscript. F.~J.~is supported by UKRI Future Leader Fellowship MR/T019956/1. This work is supported by the Deutsche Forschungsgemeinschaft (DFG) through the Research Training Group ``GRK 2149: Strong and Weak Interactions -- from Hadrons to Dark Matter'' (J.~N.). For the purpose of open access, the authors have applied a creative commons attribution (CC BY) licence to any author accepted manuscript version arising.

\appendix
\section{Irregular Monte Carlo chains \label{app:irreg}}
The treatment of error propagation and estimation within the $\Gamma$-method may be applied to observables from irregular or gapped Monte Carlo histories \cite{dobs}. This is necessary if some data has been intentionally taken on a subset of the available configurations or if data on single configurations is missing. To generalize the notation in the main text, let us consider a set of $N$ subsequent configurations $E= \{1,2,\dots,N-1,N\}$ and the subset of configurations $E_\alpha\subset E$, which has been used for the measurements of the primary observable $a_\alpha^i$. For the correct bookkeeping of configurations, we introduce the quantity
\begin{align}
    n^i_\alpha= \begin{cases}
      1\,, & \text{if}\ i \in E_\alpha \\
      0, & \text{otherwise}
    \end{cases}\,,
    \quad\text{with}\quad i\in E
\end{align}
which may be used to compute the number of configurations in $E_\alpha$ by
\begin{align}
    N_\alpha = \sum_{i=1}^N n_\alpha^i\,.
\end{align}
We now define the mean value, previously introduced in eq.~(\ref{eq:montecarlo_mean}), by
\begin{align}
    \bar{a}_\alpha=\frac{1}{N_\alpha}\sum_{i=1}^{N}n_\alpha^ia_\alpha^i\,. \label{eq:montecarlo_mean_holes}
\end{align}
and replace the fluctuations $\delta^i_\alpha$ defined on $E_\alpha$ by the fluctuations
\begin{align}
    \tilde{\delta}^i_\alpha= \begin{cases}
      \tfrac{N}{N_\alpha}(a_\alpha^i -\bar{a}_\alpha^{\phantom{i}})\,, & \text{if}\ i \in E_\alpha \\
      0, & \text{otherwise}
    \end{cases}\,,
    \quad\text{with}\quad i\in E\,,
\end{align}
where the rescaling factor $\tfrac{N}{N_\alpha}$ ensures the correct error propagation.

This change also applies to secondary observables, such that the relation
\begin{align}
\tilde{\delta}_f^i=\sum_{\alpha}\bar{f}_\alpha\tilde{\delta}_\alpha^i+\dots\,
\end{align}
generalizes eq.~(\ref{eq:delta_resampling}).
To account for the reduced number of configurations in the estimation of the autocorrelation function, we need to redefine $\Gamma_f(t)$ from eq.~(\ref{eq:gamma_function}) with an adjusted norm $\mathcal{N}_\alpha(t)$ which replaces the factor $N-t$ according to
\begin{align}
    \Gamma_f(t) = \frac{1}{\mathcal{N}_\alpha(t)}\sum_{i=1}^{N-t}\tilde{\delta}_f^i\tilde{\delta}_f^{i+t}\,,\quad \text{where}\quad
    \mathcal{N}_\alpha(t) = \sum_{i=1}^{N-t}n_f^in_f^{i+t}\,.
\end{align}
Equations~(\ref{eq:err_of_err}) and (\ref{eq:monte_carlo_error}) are then defined with the normalization factor $N_\alpha$ which gives the number of configurations that enter the analysis instead of $N$. In the limit $E_\alpha = E$ all formulae in the main text are recovered.

In terms of the implementation in \texttt{pyerrors}, eqs.~(\ref{eq:delta_fft}) and (\ref{eq:fft}) have to be adapted via
\begin{align}
(\tilde{\delta}^\prime)_f^{i} &= \begin{cases}
\tilde{\delta}_f^i & \, \,\phantom{N+}1 \leq i \leq N \\
0 & N+1 \leq i \leq N+t_\mathrm{max}
\end{cases}\,,\\
({n}^\prime)_f^{i} &= \begin{cases}
n_f^i & \, \,\phantom{N+}1 \leq i \leq N \\
0 & N+1 \leq i \leq N+t_\mathrm{max}
\end{cases}\,,\\
\Gamma_f(t) &= \frac{\Big[\operatorname{rfft}^{-1}\big(|\operatorname{rfft}\big((\tilde{\delta}^\prime)_f\big)|^2\big)\Big](t)}{\Big[\operatorname{rfft}^{-1}\big(|\operatorname{rfft}\big({(n}^\prime)_f\big)|^2\big)\Big](t)}\,,
\end{align}
to allow for the inclusion of irregular Monte Carlo chains. The rfft is employed to compute $\mathcal{N}_\alpha(t)$ efficiently.

\bibliographystyle{elsarticle-num}
\bibliography{refs.bib}

\end{document}